\documentstyle[aps,prd,epsf]{revtex}

\begin{document}

\twocolumn
[\hsize\textwidth\columnwidth\hsize\csname@twocolumnfalse\endcsname
\title{Decomposition of Hilbert space in sets of coherent states}
\author{N. Barros e S\'a\thanks{Email address: 
nunosa@vanosf.physto.se. Supported by grant PRODEP-Ac\c c\~ao 5.2.}}
\address{Fysikum, Stockholms Universitet, Box 6730, 
113 85 Stockholm, Sverige\\
{\rm and} DCTD, Universidade dos A\c cores, 
9500 Ponta Delgada, Portugal}
\date{\today}
\maketitle

\begin{abstract}
Within the generalized definition of coherent states as group orbits 
we study the orbit spaces and the orbit manifolds in the projective 
spaces constructed from linear representations. Invariant functions 
are suggested for arbitrary groups. The group $SU(2)$ is studied 
in particular and the orbit spaces of its $j=1/2$ and $j=1$ 
representations completely determined. The orbits of $SU(2)$ in 
$CP^N$ can be either $2$ or $3$ dimensional, the first of them 
being either isomorphic to $S^2$ or to $RP^2$ and the latter being 
isomorphic to quotient spaces of $RP^3$. 
We end with a look from the same perspective to the 
quantum mechanical space of states in particle mechanics. 
\end{abstract}
\pacs{02.20.Qs,03.65.Fd,42.25.Kb}
]

Coherent states are an important tool in the study of wave 
phenomena finding many relevant applications in Quantum 
physics \cite{gl1,ksu}, both in particle mechanics and in 
field theory \cite{gl3,kl4,yka,fih}. 
The familiar Glauber states \cite{schr,gl2} can be 
equivalently defined as the elements of the orbit of the 
Heisenberg-Weyl group which contains the ground state, as the eigenstates 
of the annihilation operator or as the minimum uncertainty 
wave-packets. Following these different definitions there are 
different approaches to the generalization of the concept of 
coherent states. Here we privilege the group 
theoretical approach \cite{kl1}. The generalization procedure 
has been extended to include systems with 
no classical analogue such as spin systems \cite{kl2,rad} and 
others \cite{kl3,bgi,per,bbd,ska}. For a fuller account of applications 
of coherent states in different areas of Physics see 
\cite{ksk}, where a more complete and historical list of 
references can be found. 

In the group theoretical approach to coherent states 
Hilbert space is decomposed into the union of disjoint sets of 
coherent states, the group orbits. 
In finite dimensional Hilbert spaces the orbits can be labeled 
using invariant (in the sense that they are 
constant within orbits) real functions in 
Hilbert space. 
These functions together with the group parameters 
completely parameterize Hilbert space. 
The dimensionality of the sets of coherent states can 
be related to the values these invariant functions have on the 
sets. 

Here we apply known results from group theory and invariant 
theory (reviewed in section \ref{se1} together with appendix \ref{aaaa}) 
to the study of 
coherent states as group orbits (reviewed in section \ref{se2}) 
in the complex projective spaces of Quantum mechanics (appendix \ref{aaa}). 
We make a proposal for invariant polynomial functions constructed 
from the Casimir operators in section \ref{se3}. 

The group $SU(2)$ is studied in detail in section \ref{se4}. 
Orbits turn out to be either $2$ or $3$-dimensional; the former 
are in a finite number (${\rm int}(j+1)$) within each irreducible 
representation $j$ and they are either isomorphic to $S^2$ or to 
$RP^2$; the $j=1/2$ representation is immediately solved (subsection 
\ref{s41}). In subsection \ref{s42} we work out completely the 
$j=1$ representation: the orbit space is isomorphic to a line segment; 
the orbits in its interior are isomorphic to the three-dimensional 
lens space $S^3/Z^4$ and on its vertices they are two-dimensional 
(one isomorphic to $S^2$ and the other to $RP^2$); the invariant 
function $\overline{J_i}\ \overline{J_i}$ serves as a label for the orbits. 
Our results confirm those of \cite{bac} when they overlap. 
We comment on possible approaches to the 
study of higher $j$ representations using analytical 
as well as numerical methods in subsection \ref{s43}.   
We compare our results for the $2$-dimensional orbits with the known 
formulas for coherent states in $SU(2)$ systems (subsection 
\ref{s44}). 

We finish in section \ref{se5} with the definition of invariants for the 
infinite-dimensional Hilbert spaces of particle mechanics.

\section{Group orbits and invariants}\label{se1}

Here we review the mathematical background about group orbits 
and how to label them using real functions which are invariant 
on the orbits. This 
subject can be found in the mathematical literature for Group theory 
and Invariant theory \cite{wey,bre,sch} and it has been explored in 
Physics mostly in the study of the minima of potential 
functions in theories with spontaneous symmetry breaking where these 
potentials are invariant functions in the representation space of 
the gauge group \cite{mra,mic,sla,kim,asa}. 

Let $U(g)$ be a representation of the Lie 
group $G$ with Lie algebra $\cal G$ on the manifold 
$\cal H$. We represent points in $\cal H$ by $|\psi >$, anticipating 
the application to vector spaces that we have in mind. The 
$G$-orbit through $|\phi >$ is the subset of $\cal H$ given by 
\begin{equation}
{\cal C}_\phi =\left\{ |\psi >\in {\cal H}: |\psi >=
U(g)|\phi >\ ,\ g\in G\right\}\ .\label{gor}
\end{equation}
If the group $G$ is smooth and compact, the 
$G$-orbits are smooth, closed and compact sub-manifolds of $\cal H$. 
They are also connected if $G$ is connected. 
The relation ``$|\phi '>$ lies on the same orbit as $|\phi >$'' 
is clearly an equivalence relation: 
reflexive, symmetric and transitive. As a consequence $\cal H$ can 
be partitioned into disjoint orbits 
\begin{equation}
{\cal H}=\bigcup_\phi {\cal C}_\phi 
\end{equation}
where the label $\phi$ runs over orbits (equivalence classes) and 
not over points. 
The quotient space ${\cal H}/G$ is called the 
orbit space. 

The little group (or isotropy group) of $G$ at $|\phi >$ is the 
subgroup $G_\phi$ of $G$ which leaves $|\phi >$ fixed: 
\begin{equation}
G_\phi =\left\{ g\in G: U(g)|\phi >=|\phi >\right\}\ .
\end{equation}
The subgroup 
$G_\phi$ is a Lie group and it may not be connected even if $G$ is. 
Its Lie algebra is formed by the elements of $\cal G$ which 
annihilate $|\phi >$
\begin{equation}
{\cal G}_\phi =\left\{ t\in G: t|\phi >=0\right\}\ .\label{eigo}
\end{equation}
The little groups at points lying on the same orbit are 
conjugated in $G$ and are therefore isomorphic. If 
$|\phi '>=U(g)|\phi >$ then 
\begin{equation}
G_{\phi '}=gG_\phi g^{-1}\ .
\end{equation}
The dimension of each orbit is therefore 
\begin{equation}
\dim {\cal C}_\phi =\dim G-\dim G_\phi\ .
\end{equation}
The class of all subgroups of $G$ conjugated in $G$ to $G_\phi$ 
forms an equivalence class, the orbit type $\Omega_\phi$. Distinct 
orbit types are disjoint. In the set of all orbit types a partial 
ordering relation can be introduced: 
$\Omega_{\phi '}\leq\Omega_\phi$ if an element of $\Omega_{\phi '}$ 
is conjugated to a proper subgroup of an element of $\Omega_\phi$, 
and we say that $\Omega_{\phi '}$ has a lower symmetry than 
$\Omega_\phi$. An orbit 
is said to be principal if $\Omega$ is locally minimal in orbit 
space. A point is said to be principal if it lies on a principal 
orbit. The set of all orbits with the same orbit type 
$\Omega$ is called a stratum. 

A function $f(|\psi >)$ in representation space $\cal H$ is said 
to be $G$-invariant if 
\begin{equation}
f(U(g)|\psi >)=f(|\psi >)\ ,\ \forall g\in G\ ,\ \forall |\psi >\in 
{\cal H}\ .
\end{equation}
It follows that $G$-invariant functions are also functions on orbit 
space ${\cal H}/G$. 

In appendix \ref{aaaa} we show some results and techniques 
applicable for real orthogonal linear representations (not 
necessarily irreducible) of compact groups. We are interested, 
for quantum mechanical applications, in 
complex unitary linear representations. 
But there is a standard correspondence 
between any unitary $n$-dimensional complex representation $U(g)$ of 
$G$ and an orthogonal $2n$-dimensional real representation $O(g)$, 
called the realification of $U(g)$.
In particular, since $U(N)=U(1)\times SU(N)$ all vectors in a Hilbert 
space carrying a non-trivial (in the $U(1)$ factor) representation of 
$U(N)$ which differ solely by a phase factor lie on the same orbit. 
Therefore the orbit space for the complex 
projective representations of $U(N)$ and $SU(N)$ are the same. 
For the same reason the orbit space of the complex projective 
representation of $U(N)$ is the same as the orbit space of the real 
projective representation of the realification of $U(N)$. Thus 
the orbit space of the complex projective 
representation $R$ of $SU(N)$ coincides with the projective slice 
of the realification of the representation $R\times S$ of 
$U(N)$, where $S$ is a non-trivial representation of $U(1)$. 
The orbits themselves have the same little groups and as manifolds 
they are copies of the orbits of $SU(N)$ in projective space 
multiplied by $S^1$ on account of all the 
vectors differing by a phase which are not identified in the 
latter representation. 

We finish this section with a remark about the complex 
projective spaces $P{\cal H}$ obtained after the identifications 
(\ref{proj})(see appendix \ref{aaa}). 
Unitary transformations do not change the 
norm of a vector but they may change only its phase. As a 
consequence, when using vectors $|\phi>$ in complex vector spaces 
$\cal H$ to describe points in $P{\cal H}$, the Lie algebra of the 
little group $G_\phi$ is no longer given by the elements of 
$\cal G$ which annihilate $|\phi>$ (\ref{eigo}) but rather by its 
elements for which $|\phi>$ is an eigenvector 
\begin{equation}
{\cal G}_\phi =\left\{ t\in G: t|\phi >=T|\phi>\ ,\ T\in R\right\}\ .
\label{eigi}
\end{equation}

\section{Coherent states as group orbits}\label{se2}

We follow reference \cite{kl1} and define a subset $\cal C$ 
of Hilbert space $\cal H$ to be a set of coherent states if it 
is continuous (and we represent its elements by $|c>$, $c$ 
denoting a finite number of continuous parameters) and if there 
exists a positive measure $dc$ on it 
admitting the partition of the unit operator
\begin{equation}
\int_{\cal C}|c><c|dc=1\ .
\end{equation}
Continuity guarantees that it is always possible to redefine the 
measure $dc$ in such a way that the states $|c>$ are normalized.
The existence of the partition of identity allows a functional 
representation in the parameters $c$ of vectors $\psi(c)=<c|\psi>$ 
and operators $A(c_2,c_1)=<c_2|A|c_1>$ in $\cal H$. 

For a one particle system in Mechanics the Glauber 
states can be written as 
\begin{equation}
|q,p>=U(q,p)|0>\ ,\label{rqr3}
\end{equation}
where $U(p,q)$ is the Weyl operator  
\begin{equation}
U(q,p)=e^{i(pQ-qP)/\hbar}\ .
\end{equation}
It can be shown that these are minimum 
uncertainty states since 
\begin{equation}
\Delta Q^2=\Delta P^2=\hbar /2\ ,
\end{equation}
and the equality sign is satisfied in the Heisenberg uncertainty 
relation (sometimes the square root of this relation is used; here 
we prefer this form)
\begin{equation}
\Delta Q^2\Delta P^2\leq\hbar^2/4\ .\label{hei} 
\end{equation}
These states are eigenstates of the annihilation operator and 
sometimes this property is used as there very definition. Their 
eigenvalues provide the useful analytic representation in the 
complex plane 
\begin{equation}
|p,q>=e^{(za^+-z^*a)}|0>=e^{-|z|^2/2}\sum_n\frac{z^n}{\sqrt{n!}}|n>\ ,
\label{coe}
\end{equation}
with $z=(q+ip)/\sqrt{2\hbar}$. Both the complex formulation 
(\ref{coe}) and the 
phase space formulation (\ref{rqr3}) allow for a differential 
representation of operators acting on the functions $\psi(c)$. 

The Weyl operators act as translation operators for 
position and momentum in the sense that
\begin{eqnarray}
U^+(q,p)QU(q,p)&=&Q+q\label{we1}\\
U^+(q,p)PU(q,p)&=&P+p\ ,\label{we2}
\end{eqnarray}
It follows that 
\begin{equation}
<q,p|Q|q,p>=q\ \ {\rm and}\ \ <q,p|P|q,p>=p\ .\label{wel}
\end{equation}
One can derive the properties 
\begin{eqnarray}
&&U(0,0)=1\\
&&U^{-1}(q,p)=U^+(q,p)=U(-q,-p)\\
&&U(q_2,p_2)U(q_1,p_1)=e^{i(q_1p_2-p_1q_2)/2\hbar}U(q_2+q_1,p_2+p_1)\ ,
\nonumber\\
\end{eqnarray}
which show that the Weyl operators form a group when 
acting on projective Hilbert space $P{\cal H}$ (see appendix 
\ref{aaaa}). On the whole of Hilbert 
space the Weyl operators together with an Abelian factor 
$e^{i\theta}$ form a group, the Heisenberg-Weyl group. 

Sets of generalized coherent states in particle mechanics other 
than the Glauber states fitting the definition given at 
the beginning of this section can be constructed applying the 
Weyl operators to an arbitrary vector $|\phi>$ in Hilbert 
space $\cal H$ \cite{ksk}
\begin{equation}
{\cal C}_\phi=\{ |p,q;\phi >=U(q,p)|\phi>\ , 
\ (q,p)\in R^2\} \ .\label{gcs}
\end{equation}
Like the set of Glauber states, these sets admit a differential 
representation of operators. 
But they lack the analytic representation in the 
complex plane and they are not states of minimum uncertainty 
since the vector $|\phi>$ that one starts from is arbitrary and it 
can have any values of variances $\Delta Q^2$ and $\Delta P^2$ 
{\it a priori}. They are not eigenstates of any 
particularly simple operator either.

This way of generating sets of coherent states as orbits of groups 
in Hilbert space has been generalized to representations of arbitrary 
Lie groups $G$ \cite{per}. 
Let $U(g)$, $g\in G$, be an irreducible unitary representation of $G$ 
acting on the space $\cal H$. Pick any vector $|\phi>\in\cal H$ and 
consider the $G$-orbit ${\cal C}_\phi$  
(\ref{gor}) passing through $|\phi >$. 
One can label the vectors in 
${\cal C}_\phi$ with the group elements 
\begin{equation}
{\cal C}_\phi=\left\{ |g;\phi>=U(g)|\phi>\ ,\ x\in G\right\}\ .\label{rqr}
\end{equation}
Continuity of the representation $U(g)$ 
ensures continuity of the set $\{ |g;\phi>\}$, in 
particular one has for the inner product 
\begin{eqnarray}
<g;\phi |g';\phi >&=&<\phi |U^+(g)U(g')|\phi >=\nonumber\\
&=&<\phi |U(g^{-1}g')|\phi >\ ,\label{pi}
\end{eqnarray}
which is bounded by unity. Let it exist the invariant measure $dg$ on 
the group $G$. Then if 
\begin{equation}
d=\int dg |<\phi |U(g)|\phi >|^2\label{cona}
\end{equation}
converges one has \cite{per} 
\begin{equation}
\frac{1}{d}\int dg |g;\phi><g;\phi |=1\ .\label{pid}
\end{equation}
Therefore the sets ${\cal C}_\phi$ satisfy the criteria given at 
the beginning of this section to qualify as coherent states. 
Representations obeying (\ref{cona}) are termed 
square integrable and they are always so if the volume of group 
space $\int dg$ is finite, as for compact groups. We emphasize that 
without further specification these sets of generalized coherent 
do not lead necessarily to analytic function representations \cite{ono}. 

From the definition of the orbits we see that the vectors 
$U(g)|\phi>$ for all $g$ which belongs to one left coset of the 
little group $G_\phi$ in $G$ differ from one 
another at most by a 
phase factor and that these vectors determine the same state in 
complex projective space. Thus we may label the vectors in the orbit 
${\cal C}_\phi$ with the elements $x$ of the coset space 
$X_\phi=G/G_\phi$ and we write 
\begin{equation}
{\cal C}_\phi=\left\{ |x;\phi>=U[g(x)]|\phi>\ ,\ x\in 
X_\phi\right\}\label{rqr4}
\end{equation}
where $g(x)$ is any representative $x$ of the coset. In this way we 
avoid including ``repeated'' vectors in the representation of the 
orbit as it may be the case using the set $\{ |g;\phi>\}$. In many 
cases the measure $dg$ on $G$ induces the invariant measure $dx$ 
on $X_\phi=G/G_\phi$. Then the inner product (\ref{pi}) and the 
partition of identity (\ref{pid}) become 
\begin{eqnarray}
<x;\phi |x';\phi >&=&<\phi |U[g(x)^{-1}g(x')]|\phi >\label{pii2}\\
1&=&\frac{1}{d'}\int dx |x;\phi><x;\phi |\ ,\label{pid2}
\end{eqnarray}
where 
\begin{equation}
d'=\int dx |<\phi |U[g(x)]|\phi >|^2\ .
\end{equation}
Both (\ref{pi})-(\ref{pid}) and (\ref{pii2})-(\ref{pid2}) are 
correct and it is somewhat a matter of taste which one is preferred. 
We shall use mostly the second form. 

Let us now specialize to the group $SU(2)$ which admits representations 
classified according to integer and semi-integer values $j$ with 
the Casimir operator $J^2=j(j+1)\hbar^2$. 
Let $\cal H$ be a Hilbert space carrying one such 
representation. 
The sets of coherent states (\ref{rqr}) are obtained by acting on 
any fiducial state $|\phi >\in {\cal H}$ with the group elements 
of $SU(2)$
\begin{eqnarray}
{\cal C}_\phi &=&\left\{ |\vec r>\in {\cal H}: 
|\vec r>=U(\vec r)|\phi >\ ,\ \vec r\in (4\pi )^3\right\}\label{iuu}\\
U(\vec r)&=&e^{i\vec r\cdot\vec J/\hbar}\ ,\label{rqr2}
\end{eqnarray}
where we used the so-called canonical group coordinates for 
generality. 

Using the group parameterization 
\begin{equation}
U(z,\theta)=Ne^{zJ_-/\hbar}e^{-z^*J_+/\hbar}e^{-i\theta J_z/\hbar}\ ,
\end{equation}
where $J_\pm$ are the ladder operators $J_\pm =J_x\pm iJ_y$, 
and choosing the fiducial state to be an eigenstate of $J_z$, 
$|m>$ with $m=-j,..,j$, one has \cite{rad} 
\begin{equation}
|z;m>=U(z)|z>=Ne^{zJ_-/\hbar}e^{-z^*J_+/\hbar}|m>\ ,\label{an}
\end{equation}
where the phase factor resulting from $e^{-i\theta J_z/\hbar}$ 
has been ignored (this corresponds to using (\ref{rqr4}) rather than 
(\ref{rqr})) and $N$ stands for a normalization factor. 
Further choosing $|j>$ as the fiducial state one has
$e^{-z^*J_+/\hbar}|j>=|j>$ and 
\begin{equation}
|z>=\frac{1}{(1+|z|^2)^j}e^{zJ_-}|j>\ ,\label{anal}
\end{equation}
after determination of the normalization factor. 
This analytic representation is not available in general for the 
sets (\ref{rqr2}) generated from arbitrary fiducial vectors. 

The analogous relation for spin systems to the Heisenberg 
inequality for canonically conjugate operators (\ref{hei}) is 
\begin{equation}
\Delta J_x{^2}\Delta J_y{^2}\geq \frac{\hbar^2}{4}\overline{J_z}^2\ .
\label{var4}
\end{equation}
Notice the important difference with (\ref{hei}) that now 
the right hand side of the inequality 
is not a constant. Following \cite{acs} we shall call the 
left hand side of (\ref{var4}) the uncertainty 
$\Delta J_x{^2}\Delta J_y{^2}$. 
While in particle mechanics the Glauber states saturate the Heisenberg 
inequality and they are states of minimum uncertainty, 
in spin systems the set of 
states for which the equality in (\ref{var4}) is saturated and the 
set of states of minimum uncertainty are not the same. Moreover none 
of them coincide with the set of coherent states (\ref{iuu}).

\section{Invariants for projective representations}\label{se3}

In order to construct real functions which are invariant within 
orbits (\ref{rqr4}) we make use of the inner product in Hilbert 
space. Clearly the inner product itself $<x;\phi |x;\phi >$ is such 
an invariant. It can be used to label orbits on the whole of Hilbert 
space but we are restricting attention to projective space where 
$<x;\phi |x;\phi >=1$ is a constant. Consider the generalized 
Casimir operators \cite{rac} 
\begin{equation}
C_n=c_{a_1b_1}^{b_2}c_{a_2b_2}^{b_3}...c_{a_nb_n}^{b_1}X^{a_1}X^{a_2}
...X^{a_n}\label{casi}
\end{equation}
where $c_{ab}^c$ are the structure constants of the Lie algebra 
$\cal G$ and $X_a$ its generators, 
\begin{equation}
[X_a,X_b]=c_{ab}^cX_c\ .\label{como}
\end{equation}
Indices are raised and lowered in the Lie algebra using the metric 
$g_{ab}=c_{ac}^dc_{bd}^c$. The generators of the algebra 
transform under the action of the group according to the adjoint 
representation $A_b^a(g)$
\begin{equation}
U^+(g)X^aU(g)=A_b^a(g)X^b\ .
\end{equation}
Since the Casimir operators 
commute with all generators of the algebra one has 
\begin{eqnarray}
U^+(g)C_nU(g)&=&A_{c_1}^{a_1}(g)c_{a_1b_1}^{b_2}...A_{c_n}^{a_n}(g)
c_{a_nb_n}^{b_1}\times\nonumber\\
&&\times X^{c_1}...X^{c_n}=C_n\ .\label{cas}
\end{eqnarray}
As a consequence the mean value of any Casimir operator 
$<x;\phi |C_n|x;\phi >=\overline{C_n}(x;\phi )$ is an invariant 
within orbits. But it is of no use to parameterize the orbits 
because it is actually constant within the whole irreducible 
representation. Notice however that for any polynomial in the 
generators of the algebra one has 
\begin{eqnarray}
&&\overline{X^{a_1}...X^{a_p}}(x;\phi)=<\phi |U^+[g(x)]X^{a_1}...
X^{a_p}U[g(x)]|\phi >=\nonumber\\
&&=A_{b_1}^{a_1}[g(x)]...A_{b_n}^{a_n}[g(x)]
\overline{X^{b_1}...X^{b_p}}(\phi)\ .
\end{eqnarray}
Then according to (\ref{cas}) any function of the form 
\begin{equation}
f=c_{a_1b_1}^{b_2}c_{a_2b_2}^{b_3}...c_{a_nb_n}^{b_1}
\overline{X^{a_1}X^{a_2}}\ \overline{X^{a_3}}\ 
\overline{X^{a_4}X^{a_5}X^{a_6}}...\overline{X^{a_n}}\label{cas2}
\end{equation}
where the mean values are evaluated over any combinations of the 
generators $X^a$ is an invariant within orbits. It is clear that 
using the commutator (\ref{como}) one can express any function of 
this form as a linear combination of functions of the same type 
which are real. 
To make clear what do we mean with (\ref{cas2}) let us give the 
example of the quartic Casimir operator from which the following 
invariant functions can be constructed
\begin{eqnarray}
f_1&=&c_{a_1b_1}^{b_2}c_{a_2b_2}^{b_3}c_{a_3b_3}^{b_4}
c_{a_4b_4}^{b_1}\overline{X^{a_1}X^{a_2}X^{a_3}X^{a_4}}\\
f_2&=&c_{a_1b_1}^{b_2}c_{a_2b_2}^{b_3}c_{a_3b_3}^{b_4}
c_{a_4b_4}^{b_1}\overline{X^{a_1}X^{a_2}X^{a_3}}\ 
\overline{X^{a_4}}\\
f_3&=&c_{a_1b_1}^{b_2}c_{a_2b_2}^{b_3}c_{a_3b_3}^{b_4}
c_{a_4b_4}^{b_1}\overline{X^{a_1}X^{a_2}}\ 
\overline{X^{a_3}X^{a_4}}\\
f_4&=&c_{a_1b_1}^{b_2}c_{a_2b_2}^{b_3}c_{a_3b_3}^{b_4}
c_{a_4b_4}^{b_1}\overline{X^{a_1}X^{a_2}}\ 
\overline{X^{a_3}}\ \overline{X^{a_4}}\\
f_5&=&c_{a_1b_1}^{b_2}c_{a_2b_2}^{b_3}c_{a_3b_3}^{b_4}
c_{a_4b_4}^{b_1}\overline{X^{a_1}}\ \overline{X^{a_2}}\ 
\overline{X^{a_3}}\ \overline{X^{a_4}}\ .
\end{eqnarray}
The first of this functions is the mean value of the quartic Casimir 
operator which we know to be a constant throughout all of Hilbert 
space, but there is no reason {\it a priori} why the remaining 
functions should have the same value at different orbits. On 
the other hand it is obvious that the functions $f$ of the 
generic form (\ref{cas2}) cannot all be independent in 
orbit space. At most $N$ of them can be so, $N$ being the dimension 
of orbit space. Our conjecture is that there can be found indeed 
$N$ such functions which separate the orbits in projective space 
and the values of these 
functions can then be used to parameterize the orbits.

\section{The group $SU(2)$}
\label{se4}

\subsection{General setting and the two-dimensional representation}
\label{s41}

Here we propose to study the orbit space and the invariants for the 
complex projective representations of $SU(2)$. A similar task has been 
carried out for the linear representations of $SU(2)$ in \cite{mni} and 
of $SO(3)$ in \cite{ovr}. Our problem is related to these but different, 
and it has been studied in \cite{bac}. Our presentation is complementary 
to \cite{bac} both in the methods used and in the results. 
For the projective representations of the group $SU(2)$ the element 
$g=-1$, that is the rotation by $2\pi$, always belongs to the little 
group of any vector. Therefore these representations can also be 
seen as representations of $SO(3)$. We shall for simplicity omit 
the factor $\{ 1,-1\}$ in the little groups or, which is the same, 
look upon the spaces as representations of $SO(3)$. In this section 
we take $\hbar=1$ for simplicity. We shall also consider in the 
remaining of this section that $j\ne 0$; the analysis of the 
identity representation is trivial and 
in many respects singular. 

Let $J_i\ (i=1,2,3)$ be the generators of the Lie algebra of 
$SU(2)$ with commutation relations 
\begin{equation}
[J_i,J_j]=i\epsilon_{ijk}J_k\ .\label{comu}
\end{equation}
The quadratic Casimir operator is 
\begin{equation}
J^2=J_iJ_i\ .
\end{equation}
The higher order Casimir operators in (\ref{casi}) are powers of 
$J^2$ and consequently we can think of the invariants of the type 
(\ref{cas2}) as constructed from powers of $J^2$. It is easy to 
see that up to the third power in $J^2$ all the invariants of the 
type (\ref{cas2}) can be written in terms of the following eight:
\begin{eqnarray}
f_1&=&\overline{J_i}\ \overline{J_j}\label{inv1}\\
f_2&=&\overline{J_i}\ \overline{J_j}\ \overline{J_iJ_j}\label{inv2}\\
f_3&=&\overline{J_iJ_j}\ \overline{J_jJ_i}\\
f_4&=&\overline{J_i}\ \overline{J_j}\ \overline{J_iJ_k}\ 
\overline{J_kJ_j}\\
f_5&=&\overline{J_iJ_j}\ \overline{J_jJ_k}\ 
\overline{J_kJ_i}\\
f_6&=&\overline{J_i}\ \overline{J_j}\ \overline{J_k}\ 
\overline{J_iJ_jJ_k}\\
f_7&=&\overline{J_i}\ \overline{J_jJ_k}\ \overline{J_jJ_iJ_k}\\
f_8&=&\overline{J_iJ_jJ_k}\ \overline{J_kJ_jJ_i}\ .\label{inv8}
\end{eqnarray}
All other orderings of operators can be written in terms of these 
using the commutator (\ref{comu}). These functions are real and they 
will be enough for the applications of the remaining sections. 

The Lie algebra of the little group is given by the elements 
satisfying (\ref{eigi}) 
\begin{equation}
\vec r\cdot\vec J|\psi>=\lambda|\psi>\ .\label{egii}
\end{equation}
In other words, if $|\phi>$ is not an eigenvector of angular 
momentum in some direction, then the Lie algebra of $G_\phi$ 
is trivial ${\cal G}_\phi=\{ 0\}$ and the dimension of the orbit 
${\cal C}_\phi$ is maximal, that is $\dim {\cal C}_\phi=3$ because the 
group $SU(2)$ is 3-dimensional. On the other hand, if $|\phi>$ is 
an eigenvector of angular momentum in some direction, it cannot be 
so in any other direction and the Lie algebra of its little group 
is generated by the operator of angular momentum 
$\hat r_\phi\cdot\vec J$ in that particular direction 
$\hat r_\phi$ 
for which $|\phi>$ is an eigenvector. Therefore the connected 
component of the little group $G_\phi$ is the subgroup of rotations 
around the axis in the direction $\hat r_\phi$. This is a 
$1$-dimensional 
subgroup and consequently the orbits are $2$-dimensional. 
We conclude 
that for $SU(2)$ there are only $2$ and $3$-dimensional orbits. The 
first consists of all vectors which are eigenvectors of angular 
momentum $\hat r\cdot\vec J$ in some direction $\hat r$.  
We notice that these considerations apply only to the connected 
part of the little group. There may be non-trivial discrete factors 
multiplying the connected part of the little group. In fact as we 
shall see the little group is in general not connected and orbits 
with the same dimensionality may differ in their little groups and 
therefore not be isomorphic. 

If the little group of a $3$-dimensional orbit is trivial then 
each element of $SO(3)$ defines one point in the orbit 
${\cal C}_\phi$ and ${\cal C}_\phi$ is isomorphic to $SO(3)$ which 
is in turn isomorphic to $3$-dimensional real projective space 
$RP^3$. 
If the little group is not trivial ${\cal C}_\phi$ is isomorphic to 
the coset space $SO(3)/G_\phi$ which is to say to a quotient space of 
$RP^3$ by a discrete group. 

\begin{figure}
\centerline{\epsfysize=2.8cm \epsfbox{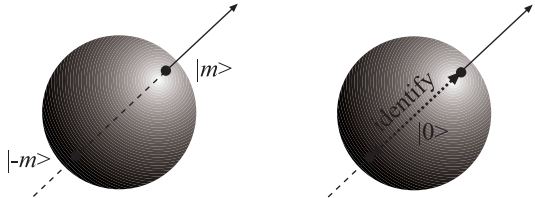}}
\caption{The two-dimensional orbits include one and only one of the 
vectors $|m>$ with $m\ge 0$. There are: $2j$ orbits isomorphic to 
$S^2$ (left) for $m\ne 0$ and if $j$ is an integer $1$ orbit 
isomorphic to $RP^2$ (right) for $m=0$.} \label{fig2}
\end{figure}

\begin{figure}
\centerline{\epsfysize=2.8cm \epsfbox{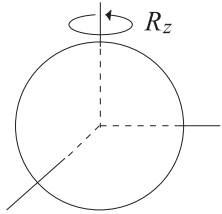}}
\caption{The little group for the $S^2$ orbits of $SU(2)$.} 
\label{fig3}
\end{figure}

\begin{figure}
\centerline{\epsfysize=6.0cm \epsfbox{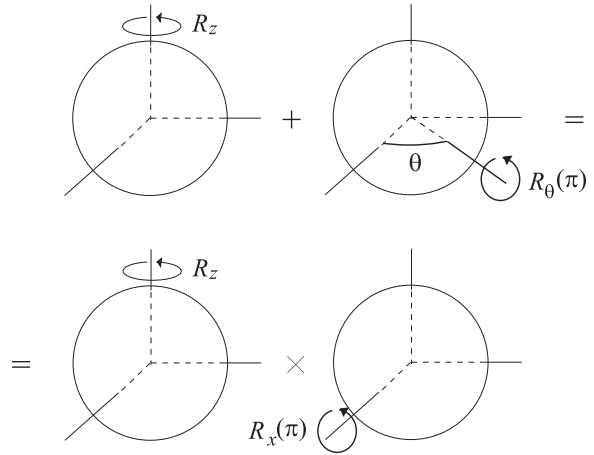}}
\caption{The little group for the $RP^2$ orbits of $SU(2)$.} 
\label{fig4}
\end{figure}

The $2$-dimensional orbits can be worked out in detail in the 
general case. We know that the eigenvalues of angular momentum in 
the $z$-direction $J_z$ are finite and non-degenerate 
\begin{equation}
J_z|m>=m|m>\ {\rm with}\ m=-j,-j+1,...,j-1,j
\end{equation}
and that 
\begin{equation}
<m|\vec J|m>=m\vec e_z\ .
\end{equation}
Applying an element of $SU(2)$ to $|m>$ clearly brings it to the 
eigenvector of the rotated direction with eigenvalue $m$. Since 
these eigenvectors are not degenerate this means that all states 
belonging to a 2-dimensional orbit can be generated after a rotation 
from one of the vectors $|m>$, or which is the same that all orbits 
contain at least one of the vectors $|m>$. It is clear also from the 
non-degeneracy of the eigenvectors that after a rotation by $\pi$ 
around any axis orthogonal to the $z$-axis the vector $|m>$ is 
mapped to $|-m>$. As a consequence for $m=0$ these rotations also 
belong to the little group of $|0>$. On the other hand they do not 
for $m\ne 0$ but one realizes that $|m>$ and $|-m>$ belong to the 
same orbit. Moreover eigenvectors in different (not parallel) 
directions cannot be identical. We conclude that 
there is a finite number of 2-dimensional orbits which can be 
generated from the vectors $|m>$ with $m\ge 0$. For $m>0$ the 
little group is the subgroup of rotations around the quantization 
axis $G_m=R_z$ and the orbit space consists of all possible 
directions which is topologically the two-sphere $S^2$. For $m=0$ the 
little group is $R_z$ plus the rotations by $\pi$ in directions 
orthogonal to the quantization axis $R_{(x,y)}(\pi)$, 
$G_0=R_z+R_{(x,y)}(\pi)=R_z\times R_x(\pi)$ and the orbit space 
consists of all possible directions (up to sign) which 
is topologically the two-dimensional projective space $RP^2$. 
The invariant (\ref{inv1}) can be used to distinguish the different 
$2$-dimensional orbits since $f_1(|m>)=m^2$. In figure \ref{fig2} 
we depict the two types of 2-dimensional orbits and in figures 
\ref{fig3} and \ref{fig4} we represent their respective little 
groups (known in the mathematical literature as $C_\infty$ and 
$D_\infty$). 

The projective space associated to the representation $j$ is 
$CP^{2j}$ and its 
real dimension is $4j$ (see the appendix). 
Its dimension is therefore greater than $2$ for $j>1/2$ and 
since the $2$-dimensional orbits are in a finite number, most of 
$CP^{2j}$ must consist of points belonging to $3$-dimensional 
orbits. Thus the dimension of orbit space is $4j-3$. For $j=1/2$ 
one has $\dim CP^1=2$ and there can be no $3$-dimensional orbits. 
On the other hand we know that there is only one $2$-dimensional 
orbit for $m=1/2$. Therefore the whole of $CP^1$ consists of 
one single $2$-dimensional orbit isomorphic to $S^2$. This is 
in agreement with the known isomorphism between $CP^1$ and $S^2$. 

We summarize this analysis of orbit space in the following three 
statements:
\medskip

\noindent\underline{I} -
{\it The orbit space of $SU(2)$ is $(4j-3)$-dimensional for its 
irreducible representations with $j>1/2$ and consists of 
$3$-dimensional orbits apart from a finite number of elements which 
are $2$-dimensional orbits. The orbit space of the representation 
$j=1/2$ consists of one single point.}
\medskip

\noindent\underline{II} -
{\it The $3$-dimensional orbits are topologically isomorphic to 
quotient spaces of $RP^3$.}
\medskip

\noindent\underline{III} -
{\it The $2$-dimensional orbits are in number of ${\rm int}(j+1)$ (integer 
part of $j+1$) and they can be distinguished by the value of the 
invariant $\overline{J_i}\ \overline{J_i}=j^2,(j-1)^2,...$ with 
minimum value $1/4$ for semi-integer $j$ representations and $0$ for 
integer $j$ representations. Topologically these orbits are 
isomorphic to 
two-spheres $S^2$ except for the $\overline{J_i}\ \overline{J_i}=0$ 
orbit of integer $j$ representations which is isomorphic to the 
two-dimensional real projective space $RP^2$.} 
\medskip

The possible little groups of the elements of the $3$-dimensional 
orbits can be found in \cite{bac}. 

\subsection{The $j=1$ representation}\label{s42}

The projective space of the representation $j=1$ 
is $4$-dimensional $CP^2$. Using the results of the previous 
subsection we can state that the orbit space is $1$-dimensional and 
consists of $3$-dimensional orbits plus two $2$-dimensional orbits, 
one isomorphic to $S^2$ ($m=1$) and the other to $RP^2$ ($m=0$). 
In terms of a $G$-invariant function $f:CP^2\to R$ labeling the 
orbits, these two $2$-dimensional orbits must lie at the vertices 
of the image of $f$ in $R$. Therefore one can state that orbit space 
is a line segment. Its interior must be of one orbit type only 
(the principal stratum) for which the orbits are some quotient 
space of $RP^3$. As a first guess for the $G$-invariant function 
to label orbits we may take (\ref{inv1}) 
$f_1=\overline{J_i}\ \overline{J_i}$. 

Now we proceed to the explicit computation of orbits using canonical 
group coordinates 
\begin{equation}
U(\vec r)=e^{i\vec r\cdot\vec\sigma}\label{ur}
\end{equation}
with
\begin{eqnarray}
&&\sigma_x=\frac{1}{\sqrt{2}}\left[ \begin{array}{ccc} 0&1&0\\ 1&0&1\\
0&1&0\end{array}\right]\ ,\ 
\sigma_y=\frac{i}{\sqrt{2}}\left[ \begin{array}{ccc} 0&-1&0\\ 
1&0&-1\\ 0&1&0\end{array}\right]\ ,\nonumber\\
&&\sigma_z=\left[ \begin{array}{ccc} 1&0&0\\ 0&0&0\\
0&0&-1\end{array}\right]\ , 
\end{eqnarray}
and the representation of $CP^2$ given by vectors of the form 
\begin{equation}
\left[\begin{array}{c} \sin\theta_1\sin\theta_2 e^{i\beta_1}\\
\cos\theta_1\\ \sin\theta_1\cos\theta_2 e^{i\beta_2}\end{array}
\right]\ .
\end{equation}
for which 
\begin{eqnarray}
f_1&=&\sin^2\theta_1\left[ \sin^2\theta_1(\cos^2\theta_2-
\sin^2\theta_2)^2+\right.\nonumber\\
&&\left. +8\cos^2\theta_1\cos\theta_2\sin\theta_2
\cos(\beta_1+\beta_2)\right] \ .
\end{eqnarray}
One has $0\le f_1\le 1$.

The eigenvalue equation (\ref{egii}) has got two families of 
solutions 
\begin{eqnarray}
|\alpha,\beta;1>&=&
\left[\begin{array}{c} \cos^2\alpha e^{-i\beta}\\
\sin (2\alpha)/\sqrt{2}\\ \sin^2\alpha e^{i\beta}
\end{array}\right]\label{orb1}\\
|\alpha,\beta;0>&=&
\left[\begin{array}{c} -\sin(2\alpha)e^{-i\beta}/\sqrt{2}\\
\cos(2\alpha)\\ \sin(2\alpha)e^{i\beta}/\sqrt{2}\end{array}
\right]\label{orb2}
\end{eqnarray}
with ranges $\alpha\in (0,\pi /2)$ and $\beta\in (0,2\pi)$. 
They represent the eigenvalues corresponding to the direction 
\begin{equation}
\hat r=(\sin 2\alpha\cos\beta,\sin 2\alpha\sin\beta,\cos 2\alpha)
\ .\label{dir}
\end{equation}
The parameter $\beta$ degenerates completely 
both at $\alpha=0$ and at $\alpha=\pi/2$. In (\ref{orb2}) 
states related by $\alpha\to\pi/2-\alpha\ ,\ \beta\to\beta+\pi$ 
correspond to the same point in $CP^2$. The first solution 
(\ref{orb1}) is the expected $S^2$ orbit and the second one 
(\ref{orb2}) is the $RP^2$ orbit. 
The vectors lying at $\alpha=0$ and $\alpha=\pi/2$ are 
easily recognizable as the eigenvectors $|1>$ and $|-1>$ 
respectively in (\ref{orb1}) and to correspond both to the 
eigenvector $|0>$ (\ref{orb2}). 

Now we check whether $f_1$ separates the orbits. 
We notice that any state belongs to the orbit of some 
state for which 
\begin{equation}
<\psi|\vec J|\psi>=J_z\vec e_z\ {\rm with}\ J_z\ge 0\ ,\label{vqu}
\end{equation}
since it is always possible to rotate a vector and bring it to 
point in the positive $z$-direction. Therefore the solution to 
(\ref{vqu}) contains at least one representative of each orbit. 
The solution to this equation consists of 
(\ref{orb2}) which we know to be composed of one single orbit plus 
the set
\begin{equation}
|\theta,\beta>=\left[\begin{array}{c} \cos\theta e^{i\beta}\\ 0\\ 
\sin\theta e^{-i\beta}\end{array}\right]\ .
\end{equation}
with $\theta\in [0,\pi/4]\ ,\ \beta\in(0,2\pi)$. But 
\begin{equation}
|\theta,\beta>=R_z(\beta)|\theta>\ {\rm with}\ |\theta>=
\left[\begin{array}{c} \cos\theta\\ 0\\ 
\sin\theta\end{array}\right]\ .\label{repr}
\end{equation}
Moreover the vector $|\theta>$ for $\theta=\pi/4$ belongs to the 
orbit (\ref{orb2}). 
Consequently among the vectors $|\theta>$ we still 
have at least one representative of each orbit. Now we compute 
\begin{equation}
f_1(|\theta>)=\cos^2\theta-\sin^2\theta=cos(2\theta)\ .
\end{equation}
Clearly the map $f_1:\theta\in [0,\pi/4]\mapsto [0,1]$ is 
one-to-one. Thus it is demonstrated that $f_1$ separates the orbits. 
The two $2$-dimensional orbits (\ref{orb1}) and (\ref{orb2}) 
lie at the extrema of 
the line segment $f_1\in [0,1]$ as predicted, 
\begin{equation}
f_1(|\alpha,\beta;0>)=0\ {\rm and}\ f_1(|\alpha,\beta;1>)=1\ .
\end{equation}

\begin{figure}
\centerline{\epsfysize=2.8cm \epsfbox{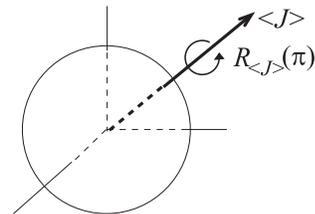}}
\caption{The little group for the $3$-dimensional orbits $S^3/Z^4$ 
of the three-dimensional representation of $SU(2)$.} \label{fig5}
\end{figure}

\begin{figure}
\centerline{\epsfysize=1.6cm \epsfbox{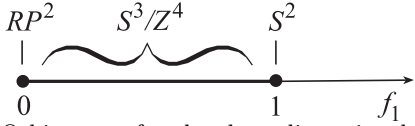}}
\caption{Orbit space for the three-dimensional representation of 
$SU(2)$.} \label{fig6}
\end{figure}

It remains to compute the little group of the orbits lying in the 
interior of $f_1\in [0,1]$. We can do it by direct calculation 
using the representatives $|\theta>$ of (\ref{repr}) and the 
explicit form of (\ref{ur}) for $j=1$ \cite{sak}
\begin{eqnarray}
U(\vec r)&=&1+\frac{i\sin r}{r}
\left[ \begin{array}{ccc} z&c^*&0\\ 
c&0&c^*\\ 0&c&-z\end{array}\right]+\nonumber\\
&&+\frac{\cos r-1}{r^2}\left[ \begin{array}{ccc} 
z^2+|c|^2&zc^*&{c^*}^2\\ zc&2|c|^2&-zc^*\\ 
c^2&-zc&z^2+|c|^2
\end{array}\right]
\end{eqnarray}
where $r^2=x^2+y^2+z^2$ and $c=(x+iy)/\sqrt{2}$. The result is 
$G_\theta=\{ 1,R_z(\pi)\}$ 
for $\theta\in ]0,\pi/4[$, that is the discrete subgroup whose 
only non-trivial element is the rotation by $\pi$ around the 
$z$-axis. By symmetry it is clear that the little group for any 
other vector $|\psi>$ belonging to a $3$-dimensional orbit is 
\begin{equation}
G_\psi=\{ 1,R_{<\psi|\vec J|\psi>}(\pi)\}\ .
\end{equation}
This is depicted in figure \ref{fig5}. 
We confirm the expectation that the 
interior of the line segment $f_1\in [0,1]$ consists of one 
single stratum of $3$-dimensional orbits. Each orbit is a lens 
space with the topology of the quotient of the 
three-sphere by the cyclic group of order $4$ \cite{llu} 
\begin{equation}
{\cal C}=RP^3/Z^2=S^3/Z^4\ .
\end{equation}
We arrived at a picture of $CP^2$ as the product of a line segment 
by $S^3/Z^4$ manifolds which degenerate to $S^2$ at one extremum 
of the segment and to $RP^2$ at the other one (figure \ref{fig6}). 

The remaining $G$-invariant functions in 
(\ref{inv2})-(\ref{inv8}) are polynomials in $f_1$ as expected 
\begin{eqnarray}
&&f_2=f_1\ ,\ f_3=2\ ,\ f_4=f_1\ ,\ f_5=2\ ,\ f_6={f_1}^2\ ,\nonumber\\ 
&&f_7=f_1\ ,\ f_8=2+f_1\ .\label{las}
\end{eqnarray}

\begin{figure}
\centerline{\epsfysize=6.0cm \epsfbox{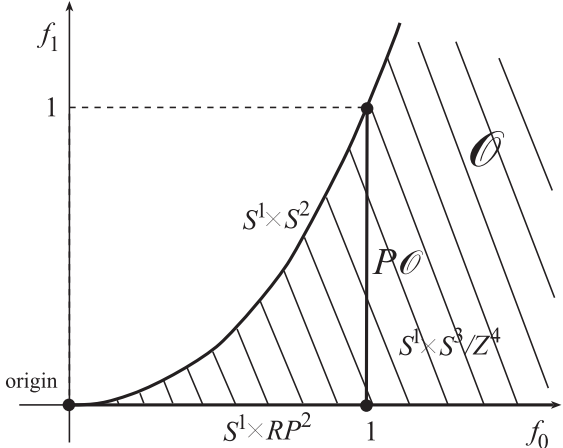}}
\caption{The orbit space for the representation $j=1$ of 
$SU(2)$ as the projective slice $P{\cal O}$ of the orbit 
space $\cal O$ of the linear representation of the 
realification of $U(2)$.} 
\label{fig7}
\end{figure}

\subsection{The $j=3/2$ representation and perspectives for future 
work}\label{s43}

In order to study the matrix $\hat P_{ij}$ of 
(\ref{matr}) we consider the whole Hilbert space of 
the representation of the realification of $U(2)$ and the two 
$G$-invariant 
functions $f_1$ and $f_0=<\psi|\psi>$ which separate the orbits. 
We have then 
\begin{eqnarray}
\hat P&=&\left[ \begin{array}{cc}\vec\nabla f_0\cdot\vec\nabla f_0&
\vec\nabla f_0\cdot\vec\nabla f_1\\
\vec\nabla f_1\cdot\vec\nabla f_0&\vec\nabla f_1\cdot\vec\nabla f_1
\end{array}\right] =\left[ \begin{array}{cc} 4f_0&8f_1\\
8f_1&16f_2\end{array}\right] =\nonumber\\
&=&\left[ \begin{array}{cc} 4f_0&8f_1\\
8f_1&16f_0f_1\end{array}\right] 
\end{eqnarray}
where the last equality is easily obtained from (\ref{las}) 
generalizing this equations to $\cal H$ by dimensional arguments. 
The values of $f_0$ and $f_1$ for which the matrix $\hat P$ is 
positive semi-definite satisfy 
\begin{equation}
f_0\ge 0\ ,\ 0\le f_1\le {f_0}^2\ .
\end{equation}
This is depicted in figure \ref{fig7}. There are 4 strata: 
the interior of this region 
is the principal stratum; the lines $\{ f_0>0,f_1={f_0}^2\}$ 
and $\{ f_0>0,f_1=0\}$ are two distinct strata composed 
respectively of $S^1\times S^2$ and $S^1\times RP^2$ orbits; and 
the point 
$\{ f_0=0,f_1=0\}$ is the $0$-dimensional stratum 
corresponding to the origin of Hilbert space. The slice 
$f_0=<\psi|\psi>=1$ gives a faithful image of orbit space in 
the projective representation. 

\begin{figure}
\centerline{\epsfysize=7.4cm \epsfbox{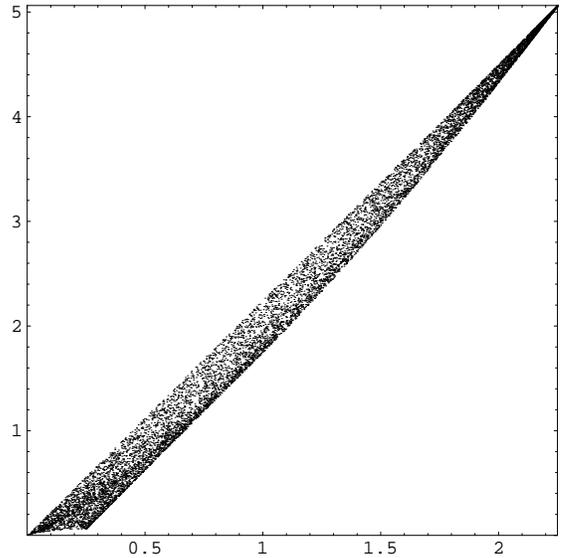}}
\caption{Numerical plot of the projection onto the plane 
$(f_1,f_2)$ of the image of orbit space for the $j=3/2$ 
representation of $SU(2)$.} 
\label{fig8}
\end{figure}

\begin{figure}
\centerline{\epsfysize=7.4cm \epsfbox{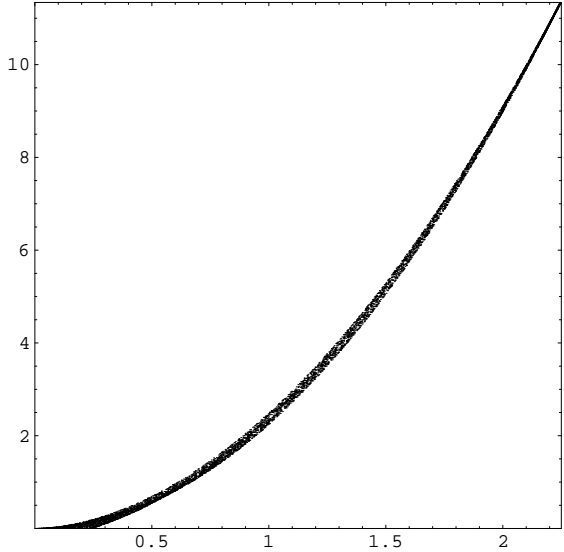}}
\caption{Numerical plot of the projection onto the plane 
$(f_1,f_8)$ of the image of orbit space for the $j=3/2$ 
representation of $SU(2)$.} 
\label{fig9}
\end{figure}

To use these techniques is one possible approach to study the higher 
dimensional representations of $SU(2)$. We also performed some 
numerical calculations on the $j=3/2$ representation. We leave 
these issues for possible future work. Here we exhibit 
in figures \ref{fig8} and \ref{fig9}, as an example, 
the numerical plots for the projections of orbit space onto the 
planes $(f_1,f_2)$ and $(f_1,f_8)$ ($f_3=f_1$ for the $j=3/2$ 
representation). This representation contains only two 
$2$-dimensional orbits isomorphic to $S^2$ according to the results 
of subsection \ref{s41} lying at the points with values of $(f_1,f_2,f_8)$: 
\begin{equation}
\left(\frac{1}{4},\frac{1}{16},\frac{1}{64}\right)\ 
{\rm and}\ \left(\frac{9}{4},\frac{81}{16},
\frac{729}{64}\right)\ .\label{popu}
\end{equation} 
In the figures one can observe the expected semi-algebraic variety 
nature of the image of orbit space. In particular one would expect 
the $2$-dimensional orbits to lie at vertices of the figures and 
indeed the kinks at the points (\ref{popu}) are visible in the 
graphics. 

\begin{figure}
\centerline{\epsfysize=3.8cm \epsfbox{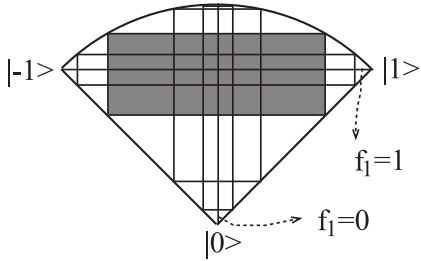}}
\caption{Orbits of the $j=1$ representation of $SU(2)$ in the 
octant picture of $CP^2$ (vertical projection).} 
\label{fig10}
\end{figure}

Numerics can also be used to study the shape of orbits in the 
picture of $CP^N$ described in 
appendix \ref{aaa}. For the octant picture of $CP^2$, figure \ref{fig12}, 
with $Z_0$ standing for the coordinate relative to the 
eigenvector $|0>$ and $Z_1$ and $Z_2$ for the coordinates relative 
to the eigenvectors $|1>$ and $|-1>$, one realizes that the vertical 
projections of the orbits form rectangles with one side parallel to 
the bisectrix of the projected quadrant. 
The bisectrix itself is a degenerate rectangle 
corresponding to the $RP^2$ orbit $f_1=0$. The other degenerate 
rectangle is the line joining the two opposed vertices of the 
quadrant and it corresponds to the $S^2$ orbit $f_1=1$. The function 
$f_1$ varies smoothly from one line to the other along the 
rectangles. The situation is depicted in figure \ref{fig10}. 

\subsection{Relation with coherent states}\label{s44}

Since the orbit space for the $j=1/2$ representation of $SU(2)$ 
consists of one 
single point, this orbit which is the whole of $CP^1$ has got 
to coincide with the set of 
coherent states (\ref{anal}) for $j=1/2$. Indeed one can work out 
explicitly (\ref{anal}) to get 
\begin{equation}
|z>=\frac{1}{\sqrt{1+|z|^2}}\left[\begin{array}{c}1\\ z\end{array}
\right]\ .
\end{equation}
The limit $z\to\infty$ defines one single point in projective 
space, meaning that the complex plane plus this point indeed 
forms a two-sphere. Setting $z=\tan\alpha e^{i\beta}$ one gets 
a standard parameterization of $CP^1$, 
\begin{equation}
|\alpha,\beta>=\left[\begin{array}{c}\cos\alpha\\ \sin\alpha 
e^{i\beta}\end{array}\right]\ ,
\end{equation}
and it is easy to check that each such vector is an eigenvector of 
$\hat r\cdot\vec J$ in the direction (\ref{dir}). 

The two orbits (\ref{orb1}) and (\ref{orb2}) of the representation 
$j=1$ are the only 
$2$-dimensional sets of coherent states of this representation and 
they must therefore coincide with the sets of 
coherent states (\ref{an}) of section \ref{se2} for $j=1$, 
whose explicit forms are 
\begin{eqnarray}
|z;1>&=&\frac{1}{1+|z|^2}\left[\begin{array}{c} 1\\ \sqrt{2}z\\ z^2
\end{array}\right]\ ,\label{ui1}\\
|z;0>&=&\frac{1}{1+4|z|^2}
\left[\begin{array}{c} \sqrt{2}z^*\\ 1\\ \sqrt{2}z\end{array}\right]
\ ,\label{ui2}\\
|z;-1>&=&\frac{1}{1+|z|^2}
\left[\begin{array}{c} {z^*}^2\\ \sqrt{2}z^*\\ 1\end{array}\right]\ .
\label{ui3}
\end{eqnarray}
The set (\ref{ui3}) coincides with (\ref{ui1}) apart from a phase $z^*/z$ 
after the redefinition $z\to 1/z^*$, which in turn coincides with 
(\ref{orb1}) for $z=\tan\alpha e^{i\beta}$. The 
set (\ref{ui2}) coincides with (\ref{orb2}) 
for $|z|=\sin\alpha/\sqrt{2\cos(2\alpha)}$ and $\arg z=\beta$.

\section{Coherent states invariants in particle mechanics}\label{se5}

The general result that Hilbert space is uniquely decomposed into 
orbits of the group generating the coherent states is still valid 
in particle mechanics. The orbits of the Heisenberg-Weyl group are the sets 
${\cal C}_\phi$ given in (\ref{gcs}). But the non-compactness of the 
Heisenberg-Weyl group and the infinite-dimensionality of Hilbert space make the 
method of finding invariants on orbits outlined in section 
\ref{se3} inapplicable. We shall therefore proceed in a different 
direction. 

From 
(\ref{we1})-(\ref{we2}) we compute 
\begin{eqnarray}
<q,p;\phi|Q|q,p;\phi>&=&<\phi|U^+(q,p)QU(q,p)|\phi>=\nonumber\\
&=&<\phi|Q|\phi>+q\\
<q,p;\phi|P|q,p;\phi>&=&<\phi|U^+(q,p)PU(q,p)|\phi>=\nonumber\\
&=&<\phi|P|\phi>+p
\ .
\end{eqnarray}
This means that within each set ${\cal C}_\phi$ all possible 
values of 
$\overline Q$ and of $\overline P$ are present. Moreover it means 
that for any two distinct vectors 
$|\phi'>,|\phi''>\in {\cal C}_\phi$ one has 
$\overline Q(\phi')\ne\overline Q(\phi'')$ or 
$\overline P(\phi')\ne\overline P(\phi'')$. Thus one can use 
$\overline Q$ and $\overline P$ as labels for the different vectors 
in ${\cal C}_\phi$. This corresponds to take as fiducial vector 
$|\phi>$ in ${\cal C}_\phi$ the unique vector for which 
$\overline Q(\phi)=\overline P(\phi)=0$. Then 
\begin{equation}
<p,q;\phi|Q|p,q;\phi>=q\ \ {\rm and}\ \ <p,q;\phi|P|p,q;\phi>=p\ ,
\label{lii}
\end{equation}
as with the Glauber states (\ref{wel}). There the vacuum 
$|0>$ is the unique vector for which $\overline Q=\overline P=0$. 
Equations (\ref{lii}) also make clear that the little group is 
trivial (the identity) everywhere in projective space. 

We notice that 
\begin{eqnarray}
U^+(q,p)(Q-\overline Q)U(q,p)&=&Q+q-\overline Q=Q\\
U^+(q,p)(P-\overline P)U(q,p)&=&P+p-\overline P=P\ .
\end{eqnarray}
Therefore the functions 
\begin{eqnarray}
&&M^{mn}=<q,p;\phi|\left\{ (Q-\overline Q)^m,(P-\overline P)^n
\right\} |q,p;\phi>=\nonumber\\
&=&<\phi|U^+(q,p)\left\{ (Q-\overline Q)^m,
(P-\overline P)^n\right\} U(q,p)|\phi>=\nonumber\\
&=&<\phi|\left\{ \left[ U^+(q,p)(Q-\overline Q)U(q,p)\right] ^m,
\right.\nonumber\\
&&\left.\left[ U^+(q,p)(P-\overline P)U(q,p)\right] ^n\right\} |\phi>=
\nonumber\\
&=&<\phi|\left\{ Q^m,P^n\right\} |\phi>\label{equ}
\end{eqnarray}
with $m$ and $n$ non-negative integers 
are invariants within ${\cal C}_\phi$. Here $\{ ,\}$ stands for the 
anti-commutator. We use it in order to make the functions $M^{mn}$ 
real since any other ordering of the operators $Q$ and $P$ in 
(\ref{equ}) can be written in terms of the $M^{mn}$ using the 
canonical commutator $[Q,P]=i\hbar$. These functions resemble 
moments of a two dimensional probability distribution, though their 
interpretation and properties are different. 

The values of $M^{mn}$ 
do not range independently over the entire real line. Besides the 
fact that for $m,n$ even one has $M^{mn}\geq 0$, the $M^{mn}$ are 
still subject to Heisenberg-like inequalities. 
These look reminiscent of the semi-algebraic variety nature of 
orbit space in the case of finite dimensional Hilbert spaces.

The relevant functions in (\ref{equ}) are actually the ones for 
which the integers $m$ and $n$ satisfy $m+n>1$ since $M^{00}=1$ is 
simply the normalization condition and $M^{01}=M^{10}=0$ by 
construction. The "second order moments" are the familiar variances 
and covariance, 
\begin{equation}
\Delta M^{20}=Q^2\ ,\ \Delta M^{02}=P^2\ ,\ M^{11}=\sigma_{QP}\ ,
\end{equation}
and the Robertson inequality (a stronger statement then the 
Heisenberg inequality \cite{sha}) reads
\begin{equation}
M^{20}M^{02}\geq\frac{1}{4}\left[ (M^{11})^2-\hbar^2\right]\ .
\end{equation}

For the Glauber states the value of the ``moments'' involved in this 
inequality is easy to compute 
\begin{equation}
M^{20}=M^{02}=\hbar /2\ ,\ M^{11}=0 
\end{equation}
confirming that they are minimum uncertainty states. It is often not 
stressed that these states not only have a minimum value for 
the uncertainty as they also have constant and identical values for 
the products involved in the uncertainty relation, the standard 
deviations of $Q$ and $P$, the same happening for all ``moments'' of 
higher order. 
For any $M^{mn}$ one can write 
the operator to be averaged $\{ Q^m,P^n\}$ in terms of the creation 
and annihilation operators $a$ and $a^+$. It is the sum of a finite 
number of monomia in $a$ and $a^+$ 
\begin{equation}
\{ Q^m,P^n\} =\sum_{i=0}^{m+n}\sum_{j=perm.}\alpha_{ij}{\cal M}_j
[a^i(a^+){^{m+n-i}}]
\end{equation}
where the index $j$ runs over the permutations ${\cal M}_j$ of 
operator ordering in $a$ and $a^+$. We have then for the Glauber 
states 
\begin{equation}
M^{mn} =\sum_{i=0}^{m+n}\sum_{j=perm.}\alpha_{ij}<0|{\cal M}_j
[a^i(a^+){^{m+n-i}}]|0>
\end{equation}
which is a finite sum of finite parcels and which is consequently 
convergent for any integer values of $m$ and $n$. 

This same argument can be used to demonstrate that all $M^{mn}$ converge 
for sets of coherent states generated 
from any eigenstate of the number operator $|\phi>=|n>$. 
And the same is true for any finite combination 
of eigenvectors of the number operator
\begin{equation}
|\phi>=\sum_{n=0}^N\alpha_n|n>\ .\label{vet}
\end{equation}
Incidentally these states seem to correspond to the "undistorted 
normalizable wave packets with classical motion" of the 
harmonic oscillator \cite{sat}.

The functions (\ref{equ}) do not 
converge on all orbits. For example, normalizability of $\psi(x)$ 
does not imply the convergence of $\int dx\ x|\psi(x)|^2$. 
But the subspace of Hilbert space where all the $M^{mn}$ 
converge is still composed of the union of entire orbits of the 
Heisenberg-Weyl group, and one may wonder whether the functions 
$M^{mn}$ separate the orbits. We leave this issue for future work. 
For the moment we notice that the $M^{mn}$ cannot separate a 
function $\psi(x)$ with an infinite degenerate zero from another 
which is identical to it on one side of the zero but which flips 
sign on the other (see the acknowledgments).

\appendix

\section{orbits in real representations}
\label{aaaa}

This appendix is taken from \cite{asa,sta} 
(sometimes literally) where the authors consider real 
finite-dimensional and orthogonal linear representations of 
compact groups. 

There is a finite number of orbit types. 
Strata are smooth disjoint sub-manifolds of $\cal H$. However they 
are not usually patched together smoothly so that the orbit space 
${\cal H}/G$ is not generally a manifold, rather 
it is a connected semi-algebraic sub-variety of $\cal H$, that is 
a subset of $\cal H$ defined by polynomial equalities and 
inequalities. The origin $|\psi >=0$ is an unique orbit 
with little group $G$, and it belongs to the maximal orbit type. 

For compact groups it can be shown that most of the orbits 
lie on a unique 
stratum of minimum orbit type called the principal 
stratum:
\medskip

\noindent 
\underline{Principal orbit theorem} - {\it The set of principal 
vectors is open and 
dense in $\cal H$; it is also connected if $G$ is connected. The set 
of principal orbits is open, dense and connected (even if $G$ is 
disconnected) in ${\cal H}/G$. All principal orbits (vectors) lie in 
a unique stratum whose orbit type is minimal in the set of 
orbit types.}
\medskip

\noindent 
From this theorem it can be shown that the boundaries of the 
principal stratum either in orbit space 
${\cal H}/G$ or in $\cal H$ are disjoint unions of the remaining 
strata which turn out to be lower-dimensional manifolds. 
The dimension of the little group is the same all over the principal 
stratum, $\dim{G_p}$, and the dimension of orbit space is given by
\begin{equation}
\dim ({\cal H}/G)=\dim {\cal H}-\dim G+\dim G_p\ .
\end{equation}

If $G$ is compact it can be shown that 
$G$-invariant functions separate the orbits, that is that for two 
distinct orbits there is at least one $G$-invariant function 
taking different values on them. The set $P_{\cal H}^G$ of all the 
real polynomials 
in $|\psi >$ (that is in its $n$ coordinates, $n$ being the 
dimension of Hilbert space) is a ring under addition and 
multiplication. An integrity basis $P_i(|\psi >)$ is a discrete 
subset of $P_{\cal H}^G$ which generates the ring $P_{\cal H}^G$ in 
the sense that any element $P\in P_{\cal H}^G$ can be written as 
\begin{equation}
P(|\psi >)=P[P_i(|\psi >)]\ .
\end{equation}
The ring of polynomial invariants $P_{\cal H}^G$ is finitely 
generated according to:
\medskip

\noindent 
\underline{Hilbert's theorem} - {\it Let $G$ be a compact Lie group 
acting orthogonally on $\cal H$. Then $P_{\cal H}^G$ admits a 
finite integrity basis.}
\medskip

\noindent 
An integrity basis can always be chosen to be minimal, in the sense 
that no proper subset of it is still an integrity basis. 
When the polynomials in the minimal integrity basis are algebraically 
independent the basis is said to be free and the representation 
$U(g)$ is said to be co-free. 

It can be 
shown that minimal integrity basis separate the orbits. This assures 
that the set of its elements can be used to parameterize the points 
in orbit space. Being $N$ the number of elements of the integrity 
basis one can think of the orbits as points in $R^N$ whose 
coordinates are the elements of the basis. The image of orbit space 
is typically not the whole $R^N$. For co-free representations 
$N=\dim {\cal H}/G$ and the image of orbit space is a subset of 
$R^N$ defined 
through inequalities between the coordinates like it happens with a 
polyhedron. 

Let $\{ P_i\}$ with $i=1,...,N$ be a minimal integrity basis and 
define the symmetric matrix 
\begin{equation}
\hat P_{ij}=\vec\nabla P_i\cdot\vec\nabla P_j\ ,\label{matr}
\end{equation}
where the inner product is performed with the same metric used for 
the inner product $<\psi|\psi'>$. Since this inner product is 
$G$-invariant, the elements of $\hat P_{ij}$ are $G$-invariant 
functions 
and according to Hilbert's theorem polynomials in the $\{ P_i\}$. 
The following important result holds:
\medskip

\noindent\underline{Theorem} - 
{\it The image of orbit space is the subset $\cal O$ of $R^N$ 
where $\hat P_{ij}$ is positive semi-definite (all its eigenvalues 
are non-negative). The subset of $\cal O$ 
where $\hat P_{ij}$ has rank $k$ is the union of all the 
$k$-dimensional strata, each of them being a connected 
component of the subset. In particular the subset of $\cal O$ 
where the rank of $\hat P_{ij}$ is maximal, that is equal to 
$\dim {\cal H}/G$, is the image of the principal stratum and 
is connected.}
\medskip

\begin{figure}
\centerline{\epsfysize=4.0cm \epsfbox{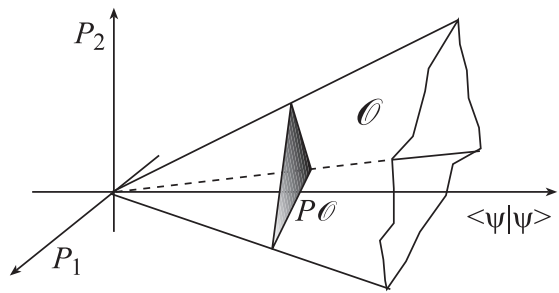}}
\caption{The orbit space for a representation on a vector space and 
the orbit space for the projective representation associated with 
the vector space.} \label{fig1}
\end{figure}

We finish with some remarks concerning projective representations, 
that is the case when one considers 
the representation space not to be the whole space $\cal H$ but 
the projective space $P{\cal H}$ of rays in $\cal H$ (see 
the appendix; here we consider $\cal H$ to be
real). 
Since $U(g)$ is linear, $G_\phi$ depends only on the direction of 
$|\phi >$
\begin{equation}
G_{\alpha|\phi>} =G_{|\phi>}\ \ {\rm for}\ \ \alpha\ne 0
\end{equation}
This means that any two vectors lying on the same ray 
have the same orbit type. Therefore the orbits in $\cal H$ 
are infinite copies along each ray of the orbits in projective space 
$P{\cal H}$ plus the origin $|\psi >=0$. For groups with no fixed 
points (apart from the origin $|\psi>=0$) the $G$-invariant 
$<\psi|\psi>\in R_+$ can always be taken 
to be one of the elements of the minimal integrity basis. Then one 
can write 
\begin{equation}
{\cal O}=P{\cal O}\times R_++\{ |0>\}
\end{equation}
where $P{\cal O}$ stands for the image of the orbit space of the 
projective representation. It turns out that most of the results of 
this section go through unchanged, 
particularly in what concerns the geometry of orbit space. 
The situation is depicted in figure \ref{fig1}. Of course the use of 
minimal integrity basis has got to be adapted. A detailed study of 
orbit spaces for projective representations can be found in 
\cite{sta}. For our purposes it suffices to mention that whenever 
necessary, such as in the application of the last theorem of this 
section one can always start with the vector space representation 
and fix $<\psi|\psi>=1$ {\it a posteriori}. 

\section{Complex projective space}\label{aaa}

Two vectors in Hilbert space $\cal H$ differing by a multiplicative 
non-zero complex constant $\alpha$ represent the same physical 
state, 
\begin{equation}
|z'>\sim |z>\ \ {\rm if}\ \ |z'>=\alpha |z>\label{proj}
\end{equation}
Therefore the space of physical states is the 
space of rays in Hilbert space or projective space $P{\cal H}$, 
that is the space of equivalence 
classes defined by (\ref{proj}) excluding the vector $|\psi>=0$. 
The projective spaces constructed from finite-dimensional Hilbert 
spaces are called $CP^N$ and are well studied spaces \cite{ben,kno}. The 
superscript $N$ stands for their complex dimension which is one unit 
lower than the complex dimension of the Hilbert space from which 
they are constructed. 

If $|n>$ is a basis for $(N+1)$-dimensional Hilbert space any vector 
$|\psi>$ can be written as
\begin{equation}
|\psi>=\sum_{n=0}^NZ_n|n>\ .
\end{equation}
The complex numbers $Z_n$ are homogeneous coordinates in $\cal H$ 
and they can also be used as coordinates in $CP^N$ provided one 
makes the identifications 
\begin{equation}
Z'_n\sim Z_n\ {\rm if}\ \exists\alpha:\forall n, Z'_n=\alpha Z_n\ .
\end{equation}

\begin{figure}
\centerline{\epsfysize=3.4cm \epsfbox{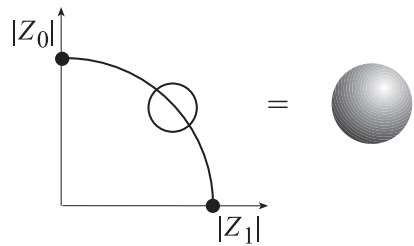}}
\caption{Complex projective space $CP^1$.} 
\label{fig11}
\end{figure}

\begin{figure}
\centerline{\epsfysize=4.2cm \epsfbox{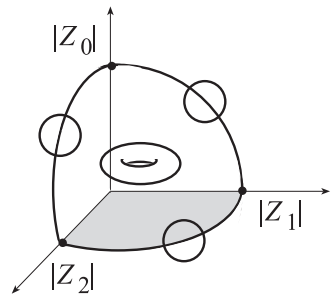}}
\caption{Complex projective space $CP^2$. The shaded region is the 
vertical projection of the octant used in figure \ref{fig8} of 
subsection \ref{s43}.} 
\label{fig12}
\end{figure}

To make a picture of how $CP^N$ looks like topologically one may 
consider the $(N+1)$-dimensional space spanned by the absolute 
values of the homogeneous coordinates $Z_i$ and set 
$\sum_{i=0}^N|Z_i|^2=1$. The resulting hyper-surface is the arch that 
bounds a quadrant for $N=1$, the curved surface of an octant for 
$N=3$, etc. These hyper-surfaces have a natural decomposition 
in smooth sets of all dimensions from $N$ down to $0$. For example 
in the case of the octant they are: the face, the three edges and 
the three vertices. 
At each point on the interior of the hyper-surfaces (that we may call 
hyper-octants) sits an $N$-torus 
because $|Z_n|\ne 0,\forall n$ and the number of relative phases is 
the maximum $N$. And on each one 
of the smooth sets mentioned before of dimension $d$ sits a 
$d$-dimensional torus because $N-d$ of the $|Z_n|$ vanish. 
In particular the vertices in this picture are points in $CP^N$ and 
not projections of tori. The lowest dimensional $CP^0$ is obviously 
nothing but a point. The situation is depicted in figures 
\ref{fig11} and \ref{fig12} for $N=1$ and $2$ 
respectively. We note that these pictures of $CP^N$ are 
more than merely topological. For example, geodesics on $CP^N$ with 
respect to the Fubini-Study metric \cite{kno} coincide in this picture 
with the ordinary geodesics on the $N$-sphere, that is, they are the 
archs of the greater circles (equators) contained in the $N$-octants. 

\bigskip
\bigskip
{\bf Acknowledgments}

I am grateful to Ingemar Bengtsson for discussions concerning 
almost all parts of this work. I thank Gerard 't Hooft for 
a comment on section \ref{se5}.


\begin{references}
\bibitem{gl1} R.Glauber, {\it Quantum optics and electronics}, eds. 
C.DeWitt, A.Blandin and C.Cohen-Tannoudji Gordon and Breach 
(New York 1964).
\bibitem{ksu} J.Klauder and E.Sudarshan, {\it Fundamentals of 
quantum optics}, Benjamin (New York 1968).
\bibitem{gl3} R.Glauber, Phys.Rev. {\bf 131} (1963) 2766.
\bibitem{kl4} J.Klauder, J.Math.Phys. {\bf 11} (1970) 609.
\bibitem{yka} Y.Ohnuki and T.Kashiwa, Prog.Theor.Phys. 
{\bf 60} (1978) 548.
\bibitem{fih} R.Field and P.Hughston, J.Math.Phys. {\bf 40} 
(1999) 2568.
\bibitem{schr} E.Schr\"odinger, p.41, {\it Collected papers on  
Wave mechanics}, Blackie and Son (London 1928).
\bibitem{gl2} R.Glauber, Phys.Rev.Lett. {\bf 10} (1963) 84.
\bibitem{kl1} J.Klauder, J.Math.Phys. {\bf 4} (1963) 1055.
\bibitem{kl2} J.Klauder, Ann.Phys.(N.Y.) {\bf 11} (1960) 123.
\bibitem{rad} J.Radcliffe, J.Phys.A:Gen.Phys. {\bf 4} (1971) 313.
\bibitem{kl3} J.Klauder, J.Math.Phys. {\bf 4} (1963) 1058.
\bibitem{bgi} A.Barut and L.Girandello, Commun.Math.Phys. 
{\bf 21} (1972) 41.
\bibitem{per} A.Peremolov, Commun.Math.Phys. {\bf 26} (1972) 222.
\bibitem{bbd} D.Bhaumik, K.Bhaumik and B.Dutta-Roy, 
J.Phys.A:Math.Gen. {\bf 9} (1976) 1507.
\bibitem{ska} B-S.Skagerstam, J.Phys.A:Math.Gen. {\bf 18} (1985) 1.
\bibitem{nie} M.Nieto, p.174, vol II of {\it Group theoretical 
methods in Physics}, Proceedings of the International seminar at 
Zvenigorod 1982, ed. M.Markov, Nauka (Moscow 1983).
\bibitem{ksk} J.Klauder and B-S.Skagerstam, {\it Coherent states - 
Applications in Physics and Mathematical physics}, World Scientific 
(Singapore 1985).
\bibitem{bac} H.Bacry, J.Math.Phys., {\bf 15} (1974) 1686.
\bibitem{wey} H.Weyl, {\it The classical groups}, 2nd ed., 
Princeton U.P. (Princeton 1946).
\bibitem{bre} G.Bredon, {\it Introduction to compact group 
transformations}, Academic Press (New York 1972).
\bibitem{sch} G.Schwarz, Invent.Math., {\bf 49} (1978) 167.
\bibitem{mra} L.Michel and L.Radicati, Ann.Phys.(N.Y.), 
{\bf 66} (1971) 758.
\bibitem{mic} L.Michel, Rev.Mod.Phys., {\bf 52} (1980) 617.
\bibitem{sla} R.Slansky, Phys.Rep., {\bf 79} (1981) 1.
\bibitem{kim} J.Kim, Nuc.Phys.B, {\bf 196} (1982) 285.
\bibitem{asa} M.Abud and G.Sartori, Ann.Phys.(N.Y.), {\bf 150} 
(1983) 307.
\bibitem{ono} E.Onofri, J.Math.Phys. {\bf 16} (1975) 1087.
\bibitem{acs} C.Aragone, E.Chalbaud and S.Salam\'o, J.Math.Phys. 
{\bf 17} (1976) 1963.
\bibitem{rac} G.Racah, {\it Group theory and spectroscopy}, 
Lecture notes-Institute for advanced study (Princeton 1951).
\bibitem{mni} J.Mickelsson and J.Niederle, Commun.Math.Phys., 
{\bf 16} (1970) 191.
\bibitem{ovr} B.Ovrut, J.Math.Phys., {\bf 19} (1978) 418.
\bibitem{llu} M.Lachi\`eze-Rey and J-P.Luminet, Phys.Rep., 
{\bf 254} (1995) 136.
\bibitem{sak} J.Sakurai, {\it Modern quantum mechanics}, 
Addison-Wesley (1994).
\bibitem{sha} R.Shankar, {\it Principles of Quantum mechanics}, 
2nd ed., Plenum Press (New York 1994).
\bibitem{sat} M.Satyanarayana, Phys.Rev.D, {\bf 32} (1985) 400.
\bibitem{sta} G.Sartori and V.Talamini, Commun.Math.Phys., {\bf 139} 
(1991) 559.
\bibitem{ben} I.Bengtsson, {\it Geometry of quantum mechanics}, 
Lecture notes (1998).
\bibitem{kno} S.Kobayashi and K.Nomizu, {\it Foundations of 
differential geometry}, Wiley (New York 1969).
\end{references}
\end{document}